\documentclass[12pt,preprint]{aastex}
\usepackage{graphicx}
\usepackage{natbib}

\shorttitle{SOFIA Observations of SN 2014J}
\shortauthors{Vacca et al.}

\begin{document}

\title{Observations of Type Ia Supernova 2014J with FLITECAM/SOFIA}

\author{William D.\ Vacca, Ryan T.\ Hamilton, Maureen Savage, Sachindev Shenoy, E.\ E.\ Becklin}
\affil{SOFIA-USRA, NASA Ames Research Center, Mail Stop N232-12, Moffet Field, CA 94035-1000}
\email{wvacca@sofia.usra.edu}

\author{and}

\author{Ian S.\ McLean, Sarah E.\ Logsdon}
\affil{Department of Physics and Astronomy, UCLA, Los Angeles, CA 90095-1562}

\author{R.\ D.\ Gehrz}
\affil{Minnesota Institute for Astrophysics, Univ.\ of Minnesota, 116 Church St. SE, Minneapolis, MN 55455-0149}

\author{J.\ Spyromilio}
\affil{European Southern Observatory, Karl-Schwarzschild-Strasse 2, Garching, Germany, D-85748}

\author{P.\ Garnavich}
\affil{Univ.\ of Notre Dame, 225 Nieuwland Science Ctr, Notre Dame, IN 46556-5670}

\author{G.\ H.\ Marion}
\affil{University of Texas at Austin, 1 University Station C1400, Austin, TX, 78712-0259}

\author{O.\ D.\ Fox}
\affil{Department of Astronomy, University of California, Berkeley, CA 94720-3411}

\begin{abstract}

We present medium resolution near-infrared (NIR) spectra, covering $1.1$ to $3.4~\mu$m, of the normal Type Ia supernova (SN Ia) SN 2014J in M82 obtained with the FLITECAM instrument aboard SOFIA
approximately $17-25$ days after maximum $B$ light.
Our $2.8-3.4~\mu$m spectra may be the first $\sim 3~\mu$m spectra of a SN Ia ever published. 
The spectra spanning the $1.5-2.7~\mu$m range 
are characterized by a strong emission feature at $\sim 1.77~\mu$m with a full width at half maximum of $\sim 11,000-13,000$ km ${\rm s}^{-1}$. 
We compare the observed FLITECAM spectra to the recent non-LTE delayed detonation models of \citet{Dessart14} and find that the models agree with the spectra remarkably well in the $1.5-2.7~\mu$m wavelength range. 
Based on this comparison we identify the $\sim 1.77~\mu$m emission peak as a blend of permitted lines of Co II.
Other features seen in the $2.0 - 2.5~\mu$m spectra are also identified as emission from permitted transitions of Co II.
However, the models are not as successful at reproducing the spectra in the $1.1 - 1.4~\mu$m range or between $2.8~\mu$m and $3.4~\mu$m. 
These observations demonstrate the promise of SOFIA 
by allowing access to wavelength regions inaccessible from the ground, and serve to draw attention to the usefulness of the regions between the standard ground-based NIR passbands for constraining SN models.


\end{abstract}


\section{Introduction}

It is widely accepted that Type Ia supernovae (SNe Ia) originate from runaway thermonuclear burning of a carbon-oxygen white dwarf that approaches or exceeds the Chandrasekhar mass limit. However, 
the evolutionary scenario (e.g., accretion from a binary companion in the so-called single-degenerate model [\citealt{Hoyle60,Whelan73}] or merger with another white dwarf in a double-degenerate model [\citealt{Webbink84,Iben84}]) that gives rise to the SN and the details of the explosion mechanisms remain uncertain.
Given the importance of SN Ia to distance determinations and cosmology \citep[see e.g.,][]{Goobar11}, it is crucial to understand the physics behind these explosions. One of the best ways to do this is to carry out 
detailed comparisons between high quality observational spectra of bright SNe Ia with physical models. Near-infrared (NIR) spectroscopic observations of SNe Ia are particularly important in this regard as they are 
not significantly affected by the presence of foreground or circumstellar dust that can seriously impact and complicate the interpretation of ultraviolet and optical spectra. In addition, the line blanketing opacity that 
dominates the UV and optical regions is lower in the NIR, making the interpretation and attribution of features easier. In order to reduce uncertainties in the model parameters and improve the accuracy of the models 
themselves, a continuous spectrum of a bright nearby SN Ia is necessary against which the models can be compared. This includes the regions between the standard ground-based NIR pass-bands that are inaccessible 
from most observatories due to strong atmospheric absorption. 
 
SN 2014J in M82 was discovered on 2014 Jan.\ 21 by \citet{Fossey14}. \citet{Zheng14} estimated the time of the explosion (`first light') to be about 7 days prior to discovery, Jan.\ 14.75 UT, while \citet{Marion14} derived the time of maximum $B$ brightness to be Feb.\ 01.74 UT. One of the nearest and brightest SNe Ia in modern history, it has been studied across a large range of wavelengths since its discovery, 
allowing the determination of a number of its physical parameters. Ultraviolet, optical, NIR, and mid-infrared observations have revealed that SN 2014J was a spectroscopically normal SN Ia, nearly identical to SN 2011fe, although with substantially higher reddening ($A_V \sim 2.0$; \citealp{Goobar14, Amanullah14, Marion14, Foley14, Zheng14}) and relatively higher velocity ($v \geq 11,000$ km s$^{-1}$) absorption features in its optical and NIR spectra \citep{Goobar14, Marion14}. The analysis 
of the optical and NIR spectra presented by \citet{Marion14} reveals radial stratification of elements produced during the explosion, indicative of relatively little mixing and consistent with that expected by delayed detonation models. \citet{Telesco14} were able to reproduce successfully the $8-13~\mu$m spectra of SN 2014J, which are dominated by strong lines of [Co II] and [Co III], with a model from \citet{Hoeflich02} for a spherical delayed detonation explosion on a Chandrasekhar-mass white dwarf. Based on fits to the observed gamma-ray line fluxes, \citet{Churazov14} and \citet{Diehl15} estimated that $\sim 0.5-0.6~M_\odot$ of $^{56}$Ni was produced in the explosion, in agreement with the model used by \citet{Telesco14}.
 
Because SN 2014J occurred in a nearby galaxy ($D \sim 3.4$ Mpc; \citealp{Dalcanton09}) and therefore was relatively bright, it constitutes a nearly ideal object on which to test SN Ia spectral models. Fortuitously, it also 
provided an opportunity to test the FLITECAM instrument \citep{McLean06} on the NASA Stratospheric Observatory for Infrared Astronomy (SOFIA), which can be used to fill in some of the wavelength gaps in NIR spectra of 
SNe Ia obtained from ground-based observations. NIR observations obtained with SOFIA are much less susceptible to the deleterious effects of atmospheric water vapor, whose strong absorptions serve to define the standard 
ground-based $J$ ($\sim 1.1 - 1.35~\mu$m), $H$ ($\sim 1.5 - 1.75~\mu$m), and $K$ ($\sim 2.0 - 2.35~\mu$m) bands. 
At altitudes $>39,000$ feet ($\sim 12$ km), typical of SOFIA flights, the precipitable water vapor overburden is generally between a few to tens of microns whereas ground-based water vapor overburdens are typically on the 
order of a few millimeters, even at the highest and driest sites \citep[e.g.,][]{Giovanelli01,Radford08}. A description of SOFIA 
and its instrument suite can be found in the papers by \citet{Young12}, \citet{Gehrz09}, and \citet{Becklin07}. In this paper we present the first spectra obtained by FLITECAM during its commissioning flights on board SOFIA. 
The NIR spectra of SN 2014J obtained with FLITECAM/SOFIA and discussed here comprise a complement to the data sets presented by \citet{Marion14} and \citet{Friesen14}.

\section{Observations and Data Reduction}
SN 2014J was observed with FLITECAM on four separate SOFIA flights (flight nos.\ 146, 147, 148, and 149) on 2014 Feb 19, 21, 25, and 27 UT, approximately 17-25 days after the time of maximum $B$ brightness, 
or approximately 36-44 days after first light \citep{Zheng14}. The flights originated in Palmdale, CA and the primary goal was the commissioning of the FLITECAM instrument on the SOFIA telescope.
FLITECAM is a NIR imager and grism spectrograph covering the $1 - 5$ micron range \citep{McLean06, Smith08, Logsdon14} and both images (see Fig. 1) and spectra of the SN were obtained. The observations were
acquired at four different altitudes between 38,000 and 43,000 feet and a range of air masses. The full-width at half maximum (FWHM) of the images and spectra on these flights was $\sim 3\arcsec$.
Dithered images were obtained in the FLITECAM $J$, $H$, and $K$ filters.\footnote{The FLITECAM filter passbands are available at \\
{\tt http://www.sofia.usra.edu/Science/ObserversHandbook/FLITECAM.html}.} 
FLITECAM was co-mounted with the HIPO instrument during these observations, a configuration that precluded observations at wavelengths longwards of $\sim 4~\mu$m, and reduced the sensitivity longwards of 
$\sim 2~\mu$m, due to high background levels resulting from the warm dichroic and transfer optics. 
The images were reduced in a 
manner typical for NIR images.
Flat fields and background sky frames were constructed from dithered observations obtained $6\arcmin$ away from the target. Individual frames were flat-fielded, sky-subtracted, and then shifted and median 
combined to produce a final image in each filter. The frames in each filter were then aligned and combined. Photometric calibrations of the images were determined from measurements of 
the star BD+70 587, which is present on the FLITECAM images of SN 2014J and for which 2MASS magnitudes are available. The resulting SN photometry is presented in Table 1. 
The agreement with the values given by \citet{Amanullah14}, \citet{Marion14}, and \citet{Foley14} for times near those of our SOFIA observations is reasonably good.\footnote{
We note that, because the spectrum of SN 2014J is unlike that for any standard star, the effective wavelengths of the filters and therefore the color corrections could be substantial, 
particularly in the $J$ and $H$ bands. However, in order to facilitate comparisons with magnitudes of the SN given in the literature, we have not made any such corrections to the values given in Table 1.}

The spectra were obtained by nodding the target between two different positions along the slit, referred to as A and B respectively.  Due to the fast rotation of the field and the constraints of pointing the telescope, 
the spectra were acquired via several sequences of AB observations.
The slit is $60\arcsec$ long and the A and B positions are separated by $30''$. The low-resolution wide 
slit was used, which resulted in a spectral resolving power R of $\sim 1300$.  The log of the spectroscopic observations
is given in Table 2. The data were reduced using {\tt Redux}, the SOFIA facility pipeline \citep{Clarke14} incorporating the {\tt fspextool} software package, which is modification of the Spextool package \citep{Cushing04} developed
for the SpeX NIR spectrograph at the NASA Infrared Telescope Facility \citep{Rayner03}.  {\tt fspextool} performs non-linearity correction, AB pair subtraction, source profile construction, extraction and background aperture definition, optimal extraction, 
and wavelength calibration for FLITECAM grism data. Wavelength calibration was performed using the OH emission lines in the spectra of the background sky.

In order to account for the differences in the detector response at the two positions (A and B) along the slit, we did not combine the A and B spectra until after the telluric
correction and flux calibration steps. The latter steps were carried out using the {\tt xtellcor\_general} package \citep{Vacca03}, and for these purposes we observed an A0V star during each flight. The spectra 
of the A0V star were reduced in exactly the same way as for the SN. Again, we kept the A and B beam spectra separate. We flux calibrated the A beam spectra of the SN with the A beam spectra of the standard
and did the same for the B beam spectra. To remove residual telluric features resulting from the differences in observing altitude and air mass between the standard star and the SN, we computed telluric absorption 
models appropriate for the observing conditions, using the ATRAN code \citep{Lord92}, smoothed and binned them to the FLITECAM resolution and sampling, and generated correction curves as a function of 
wavelength from the ratios of the telluric models for the standard to those for the SN. These corrections were then applied to the spectra of the SN. The telluric-corrected and flux-calibrated A and B beam spectra of the 
SN were then combined. The spectra from each grism setting for a given date were then merged to produce a final spectrum spanning the full wavelength range available. We then scaled the spectra so that 
synthetic photometry yielded values that matched the imaging photometry values provided by both our own images and those of \citet{Foley14}, \citet{Amanullah14}, and \citet{Marion14} interpolated to the dates of our observations. 

\section{Results}
The extracted FLITECAM spectra of SN 2014J are shown in Fig.\ 2. Wavelength regions that are inaccessible, or difficult to observe, from the ground due to strong telluric absorption are marked. 
In Fig.\ 3 we present a comparison between the spectra from Feb.\ 19 and Feb.\ 27 over the $1.5-2.7~\mu$m wavelength range. The signal-to-noise ratio (S/N) for these spectra ranges from $\sim 10$ to 
more than $60$ per pixel at the highest flux levels; over most of the wavelength range, the S/N is $> 20$ per pixel.

The most striking aspects of the spectra shown in Figs.\ 2 and 3 are the strong emission features at $1.55-1.65~\mu$m and $\sim 1.77~\mu$m. The former emission feature can clearly be seen in the 
later-time ground-based spectra of SN 2014J presented by \citet{Marion14}. Although the latter feature is present in ground-based spectra of SN 2014J \citep{Marion14}, and can also be seen in some of the 
spectra of other SN Ia presented by \citet{Gall12}, \citet{Hsiao13}, and \citet{Marion09,Marion03}, the FLITECAM spectra are the only data that span the entire feature profile; the ground-based 
spectra cover only a portion of this wavelength region due to the poor and variable atmospheric transmission between $\sim 1.8$ and $\sim 2.0$ microns. The shapes and peak wavelengths of both features 
vary substantially between the two sets of observations, which were obtained only $\sim 8$ days apart. A Gaussian fit to the $1.77~\mu$m feature in the Feb.\ 19 spectrum yields a centroid wavelength of 
$1.768~\mu$m and a full width at half maximum of $\sim 0.080~ \mu$m ($\sim 13,500$ km s$^{-1}$); on Feb.\ 27, the feature is centered at $1.784~ \mu$m with a width of $\sim 0.062~ \mu$m ($\sim 10,500$ km s$^{-1}$). 
Uncertainties are on the order of $0.003~\mu$m ($\sim 500$ km s$^{-1}$). If the observed wavelength shift were to reflect a change in 
velocity of a specific emission feature, this would correspond to a velocity redshift of about 2,700 km s$^{-1}$, with a decrease in the velocity width of a similar amount. Although the ground-based data presented by 
\citet{Marion14} confirm the shift seen in the FLITECAM spectra, the full wavelength coverage provided by FLITECAM in the $1.8-2.0~\mu$m region allows the change in the profile to be identified easily. A 
shift in the feature wavelength is not readily apparent in the NIR spectra of other SNe Ia shown by \citet{Marion09} but can be discerned in the spectra of the normal SN Ia SN2011fe presented by \citet{Hsiao13}.

Based on the model of \citet{Wheeler98}, \citet{Marion03, Marion09} identify the emission in the $1.7~\mu$m region as arising from a blend of Fe group lines along with absorption from Co II. 
However, the strength and symmetric Gaussian shape of the $\sim 1.77~\mu$m feature argue against this interpretation and in favor of (primarily) a single emission feature. \citet{Dessart14} identify this 
feature, as well as other strong features seen in the $K$ band spectra of SNe Ia, as emission from permitted lines of Co II. In fact, the strengthening of the Co lines is responsible for the well-known 
secondary peak in the broadband NIR photometric light curves of SNe Ia in their model. Their DDC10 A4D1 (hereafter DDC10) model spectra 
exhibit a strong Co II emission feature at $\sim 1.75~\mu$m, arising from a blend of lines, primarily those at $1.7463$, $1.7770$ and $1.8069~\mu$m, starting at about 25 days after explosion 
and shifting redward with time. The redshift is the result of the variation in line strengths of the various components. 
Similarly, \citet{Gall12} identify both the $1.55~\mu$m and the $1.77~\mu$m feature seen
in the NIR spectra of the normal SN Ia SN 2005cf and three other SNe Ia at 9-12 days past maximum $B$ band emission, as permitted lines of Co II, based on comparisons with the W7 model of \citet{Nomoto84}. 

In Fig.\ 3 we compare our combined $1.5-2.7~\mu$m spectra of SN 2014J at 18 and 26 days past $B$ maximum (36 and 44 days past explosion) with the DDC10 model spectra\footnote{
Kindly provided to us by L.\ Dessart and S.\ Blondin},
computed for dates close to those of our FLITECAM observations. The DDC10 A4D1 model is a set of one-dimensional (i.e., spherically symmetric), time-dependent, non-LTE calculations for a delayed detonation 
explosion of a Chandrasekhar-mass white dwarf at a distance of 10 pc. The model includes non-thermal effects, non-local energy deposition, ``huge" model atoms for iron group elements, 
and forbidden lines of both intermediate mass elements and iron group elements; descriptions of the model can be found in  \citet{Dessart14}, \citet{Blondin13}, and references therein. We shifted the model 
spectra to account for the redshift of M82 ($v = 203$ km s$^{-1}$).
We also reddened the model spectra to account for both foreground reddening, due to both the Milky Way and M82, and circumstellar reddening and scattering, using the extinction values and the power law reddening/scattering curve advocated by \citet{Amanullah14} (see also \citealp{Foley14}). 
We then scaled the model spectra to match the overall flux levels of the observed data; a single scale factor for each model spectrum was derived from a least-squares fit to the observed spectra.
The strongest emission lines of Co II and [Co III], as predicted by the model \citep{Blondin15}, are identified near the bottom of each plot. (The wavelengths of NIR Co II lines can be found in the tables given by \citealp{Marion09} and \citealp{Gall12}.) 
As can be seen, the model spectra match the observed spectra extremely well in this wavelength regime, although the time past explosion for the best-fitting model to the earlier of the two observed spectra 
is somewhat discrepant. 
For an explosion date of 2014 Jan.\ 14.75 UT \citep{Zheng14}, the FLITECAM spectra obtained on Feb.\ 19 correspond to $\sim 36$ days after first light, and the spectra obtained on Feb.\ 27 correspond to 
$\sim 44$ days past first light. The model spectrum that matches the earlier data set best corresponds to $33.15$ days past explosion. For the later data set, the model spectrum corresponding to $40.11$ days past explosion 
provides the formal best fit to the data; however, the model spectra for $44.12$ and $48.53$ days past explosion fit the observed data in the region around the $1.77~\mu$m feature considerably better.

Since the model spectra are computed for a distance of 10 pc, the scale factors needed to match the model spectra to the observed flux levels yield estimates of the distance to SN2014J and 
therefore provide additional constraints on the models, and the predicted absolute fluxes, that are independent of the overall spectral shape. Further, the distances derived from scaling the model spectra to 
observed spectra on different dates and over different wavelength ranges provide a test of the self-consistency of the models.
We find that the distances derived from the scale factors are well
within the range of values for M82 ($D = 3.2 - 5.5$ Mpc with most values between $3.2-3.9$ Mpc; \citealp[for example][]{Kara06,Dalcanton09,Tutui97}). The best fit reddened model spectra shown in Fig.\ 3 yield distances of 3.7 Mpc (for the model corresponding to 33.15 days past explosion) and 4.0 Mpc (for the model corresponding to 44.12 days past explosion); the model spectra for $40.11$ and $48.53$ days past 
explosion yield distances of 4.2 Mpc and 3.5 Mpc, respectively. The agreement of the values, both with each other and independent distance estimates, is fairly good, especially in light of the ($\sim 5-10$\%) uncertainties in the absolute flux levels of the observed spectra resulting from the scaling needed to match the interpolated photometric values. Finally, the comparison with the models also serves to illustrate the relative insensitivity of NIR observations to the effects of reddening, even for an object such as SN 2014J for which $A_V \sim 2.0$ mag \citep{Amanullah14, Foley14}. 

The remarkably good agreement between the observed spectra and the DDC10 model spectra indicates that this model captures the basic elements of the physics involved in the formation and evolution of the observed NIR spectra of SNe Ia. 
The comparison between the model spectra and the observed spectra in the $H$ and $K$ band further supports the conclusions of \citet{Gall12} and \citet{Friesen14}, who suggested that the post-maximum spectra of 
normal SNe Ia, including that of SN 2014J, longward of $\sim 1.5~\mu$m are dominated by broad emission from permitted lines of Co II, but calls into question the previous identification of Co II absorption features 
at high velocity in the $K$ band spectra of other SNe Ia by \citet{Marion09}.
It also suggests that the identification of the $1.77~\mu$m feature as emission from Co II, dominated by the $1.7770~\mu$m transition, is probably correct, despite the fact that the best fitting model spectrum predicts 
a somewhat wider and shifted line compared to the observations on Feb.\ 19. The discrepancy between the observed and predicted line shape suggests that the relative strengths of the Co II lines contributing to the $1.77~\mu$m feature are incorrect in the model. The $1.7463~\mu$m line is far stronger in the model than the observations on this date indicate.
As demonstrated in Fig.\ 4, fitting the emission feature with multiple Gaussians, constrained to have the same velocity width and centroids fixed at  $1.746$, $1.777$ and $1.807~\mu$m, reveals that the observed 
line shape in the data from Feb.\ 19 can be successfully reproduced with relative line intensities in the ratio of $\sim 0.45 : 1 : 0$. To reproduce the line shape observed in the Feb.\ 27 data requires of ratios of relative intensities of  $\sim 0.25 : 1 : 0.65$. Therefore, the observed shift in the $1.77~\mu$m feature is due to the weakening of the $1.7463~\mu$m line and the strengthening of the $1.8069~\mu$m line. The line
widths also decrease from $\sim 12,000$ km s$^{-1}$ to $\sim 7,000$ km s$^{-1}$ as the ejecta expand and the optical depth decreases.

Following the suggestion of \citet{Gall12}, we used the width of the unblended Co II lines at $2.36~\mu$m and $2.46~\mu$m to derive an estimate 
for the extent in velocity space of the Co-rich core. The weaker Co II line at $2.36~\mu$m seen in the Feb.\ 27 data yields a velocity extent of $\sim 7,000$ km s$^{-1}$, in 
agreement with the line widths estimated above for the components of the $1.77~\mu$m feature. The stronger Co II $2.46~\mu$m line, the full width of which 
is generally inaccessible from the ground, yields a velocity extent of the Co core of $\sim 11,000$ km s$^{-1}$ for both observation dates (Feb.\ 19 and Feb.\ 27),
in good agreement with the location of the iron-group zone in models such as W7 and DDC10. Since almost all of the Co results from the decay of $^{56}$Ni, the velocity extent of the Co should reflect the distribution
of $^{56}$Ni immediately after the explosion. As can be seen in Fig.\ 1 of \citet{Dessart14}, the DDC10 model predicts that 99\% of the $^{56}$Ni should be located within $\sim 11,600$ km s$^{-1}$ \citep{Blondin13, Blondin15}. Our estimated velocity range is in good agreement with this value, as well as with the values found by \citet{Gall12} for SN 2005cf and SN 2002bo at $\sim 30-40$ days past maximum $B$ light
(slightly later than our observations of SN 2014J at $\sim 18-26$ days past maximum $B$ light).

Despite the success of the DDC10 model in reproducing the $H$ and $K$ band spectra of SN 2014J, the model does not match as well at other wavelengths. In Fig.\ 5 we show the comparison between the 
FLITECAM spectra obtained on Feb.\ 25 and 27 and the DDC10 model over the entire $1-3.5~\mu$m range. Clear discrepancies between the data and the model are readily apparent in the $J$ band, the region around 
$2 ~\mu$m, and longwards of $\sim 3 ~\mu$m. The disagreement between the model spectra and the observed spectrum in the $2~\mu$m region at 44 days past explosion is somewhat surprising given how well the model matches the spectrum in this wavelength region earlier in the temporal evolution. Although \citet{Friesen14} suggest that the $2~\mu$m feature seen in their spectra of a sample of SNe Ia (including SN 2014J) is due to emission from [Ni II] at $1.939~\mu$m, the DDC10 model indicates that this feature 
is due primarily to [Co III] at $2.002~\mu$m. In fact, very little $^{56}$Ni should remain at this time,
as 44 days past explosion is $\sim 7$ times longer than the 6 day half-life of $^{56}$Ni, and therefore \citet{Friesen14} speculated that the [Ni II] emission might arise from the stable isotope $^{58}$Ni.
However, the observed emission feature exhibits a peak at $\sim 1.99~\mu$m, a wavelength which is far longer than that expected if [Ni II] were the primary component ($\sim 1.94~\mu$m).
\citet{Friesen14} noted a similar wavelength discrepancy for this feature (relative to the predicted emission from [Ni II]) but attributed the wavelength difference to a redshift of the observed emission due to the combined effects of an asymmetrical explosion and Doppler shifts. Attributing the emission to [Co III] instead seems to be a far more satisfying 
explanation, especially given the success that spherical models have in reproducing the spectra. Nevertheless, the DDC10 model predicts this [Co III] feature decreases in strength far faster than the observations 
indicate; the emission peak is still very strong $44$ days past explosion, whereas the models predict the feature should have faded completely by about day 40.

In Fig.\ 6 we present a comparison between the DDC10 model spectra and our $J$ band spectra. While the model seems to agree reasonably well with our observed $J$ band spectra at the earlier time in the evolution ($\sim 36$ days past explosion), the mismatch between the observations and the model at later times ($\sim 44$ days past explosion) is particularly bad. The most prominent characteristic of this wavelength region is a relatively strong emission peak at $\sim 1.29~\mu$m, present in both of our $J$ band spectra (Fig.\ 1). 
The model spectra, particularly for late times in the temporal evolution of the SN, do not adequately reproduce this feature.
Similar emission peaks can be seen in some of the late time spectra presented by \citet{Marion09} and \citet{Gall12}, who attribute the emission to a blend of Fe II lines. Again, as for the $1.77~\mu$m and $2.46~\mu$m features, the significantly higher atmospheric transmission afforded by SOFIA observations allows this $1.29~\mu$m feature to be discerned and measured easily in the FLITECAM spectra. The comparison between the observed $2.8-3.4~\mu$m spectrum and the model spectra is shown in Fig.\ 7, where a large discrepancy is also apparent, despite the low signal-to-noise ratio of the data. 
(The low signal-to-noise of this spectrum is partially due to the additional background from the warm transfer optics in the dual instrument configuration.) 
The FLITECAM data indicate that broad emission is present between $\sim 3.0~\mu$m and $3.4~\mu$m, which can provide additional constraints on models of SNe Ia. This spectral region is very difficult to observe from the ground due to strong telluric absorption, and our data may represent the first $\sim 3~\mu$m spectra of a SN Ia ever obtained.

\section{Conclusions}

By providing access to wavelength regions that are difficult or impossible to observe from the ground, SOFIA can contribute significantly to studies of nearby bright SNe. We have demonstrated this with our FLITECAM
observations of a normal SN Ia, SN 2014J in M82, about $36-44$ days after explosion, which have fully revealed a strong emission feature at $1.77~\mu$m. This feature shifted redward in wavelength and simultaneously narrowed between our two sets of observations separated by $\sim 8$ days. Because this feature extends into a wavelength regime that is difficult to observe from the ground, the identification, width, and the shift of the feature have never before been appreciated. Comparison with the non-LTE delayed detonation DDC10 A4D1 model of \citet{Dessart14} suggests that this feature is most likely due to a blend of permitted emission lines from Co II. The variation in location and width is most likely due to changes in the relative strengths of the component lines.
Other peaks seen in the $K$ band spectra are also due to permitted Co II lines emitting at the systemic velocity, as opposed to highly redshifted absorption features as suggested previously. The width of the Co II $2.46~\mu$m, a transition which is generally inaccessible from the ground, indicates that the Co-rich core extends to about 11,000 km s$^{-1}$ in agreement with predictions from models. The \citet{Dessart14} model matches the observed spectra remarkably well across the $1.5-2.6~ \mu$m range, and scaling the model spectra to match the observed flux levels yields distances consistent with previous estimates for M82. However, the model provides a much poorer match to the observed spectra in the $J$ band, and at $2~\mu$m and $\sim 3~\mu$m around $44$ days after explosion ($\sim 26$ days after $B$ maximum). The $J$ band spectra 
are characterized by a strong broad emission peak at $\sim 1.29~\mu$m, which is not reproduced in the model, particularly at late times in the temporal evolution . Similarly, the observed spectrum between $3.0~\mu$m and $3.4~\mu$m, the first SN Ia spectrum ever obtained in this wavelength range, exhibits broad emission that is not seen in the model. The region around $2~\mu$m exhibits a strong emission feature of [Co III] that decreases in strength far less quickly than the model predicts. Our results serve to draw attention to the usefulness of the NIR regime, including the regions between the standard ground-based NIR passbands, for constraining SN models, due in part to the relative insensitivity of the NIR to the effects of reddening. The wavelength coverage afforded by FLITECAM on SOFIA  allows the identification and study of the evolution of features not previously recognized in ground-based spectra of SNe Ia.

\acknowledgments

We would like to thank the USRA/DSI science and mission operations teams and the engineering and support staff at NASA Armstrong and Ames for their pivotal roles in making SOFIA a reality. We also thank Luc Dessart and St{\'e}phane Blondin for sharing their SN models with us. RDG thanks R.\ G.\ Arendt, N.\ M.\ Ashok, D.\ P.\ K.\ Banerjee, E.\ Dwek, A.\ Evans, M.\ Greenhouse, D.\ Shenoy, S.\ Starrfield, T.\ Temim, D.\ H.\ Wooden,
and C.\ E.\ Woodward for providing useful input during the formulation of the scientific case for the observing program as members of the DDT proposal team.
Similarly, PG thanks R.\ Kirshner, P.\ Milne, E.\ Hsiao, M.\ Phillips, and N.\ Suntzeff for their contributions to the science case of the accepted DDT proposal.

Based on observations made with the NASA/DLR Stratospheric Observatory for Infrared Astronomy (SOFIA). SOFIA is jointly operated by the Universities Space Research Association, Inc. (USRA), under NASA contract NAS2-97001, and the Deutsches SOFIA Institut (DSI) under DLR contract 50 OK 0901 to the University of Stuttgart. Financial support for RTH, RDG, and PG was provided in part by SOFIA Cycle 2 GI Research Grant \#'s 75-0001, 75-0002, and 02-0100, respectively, issued by USRA, on behalf of NASA. 
ISM and SEL were supported by NASA through grant 08500-05 from USRA for the development of FLITECAM.

\newpage

\begin{figure}
\begin{center}
\includegraphics[height=4.0in]{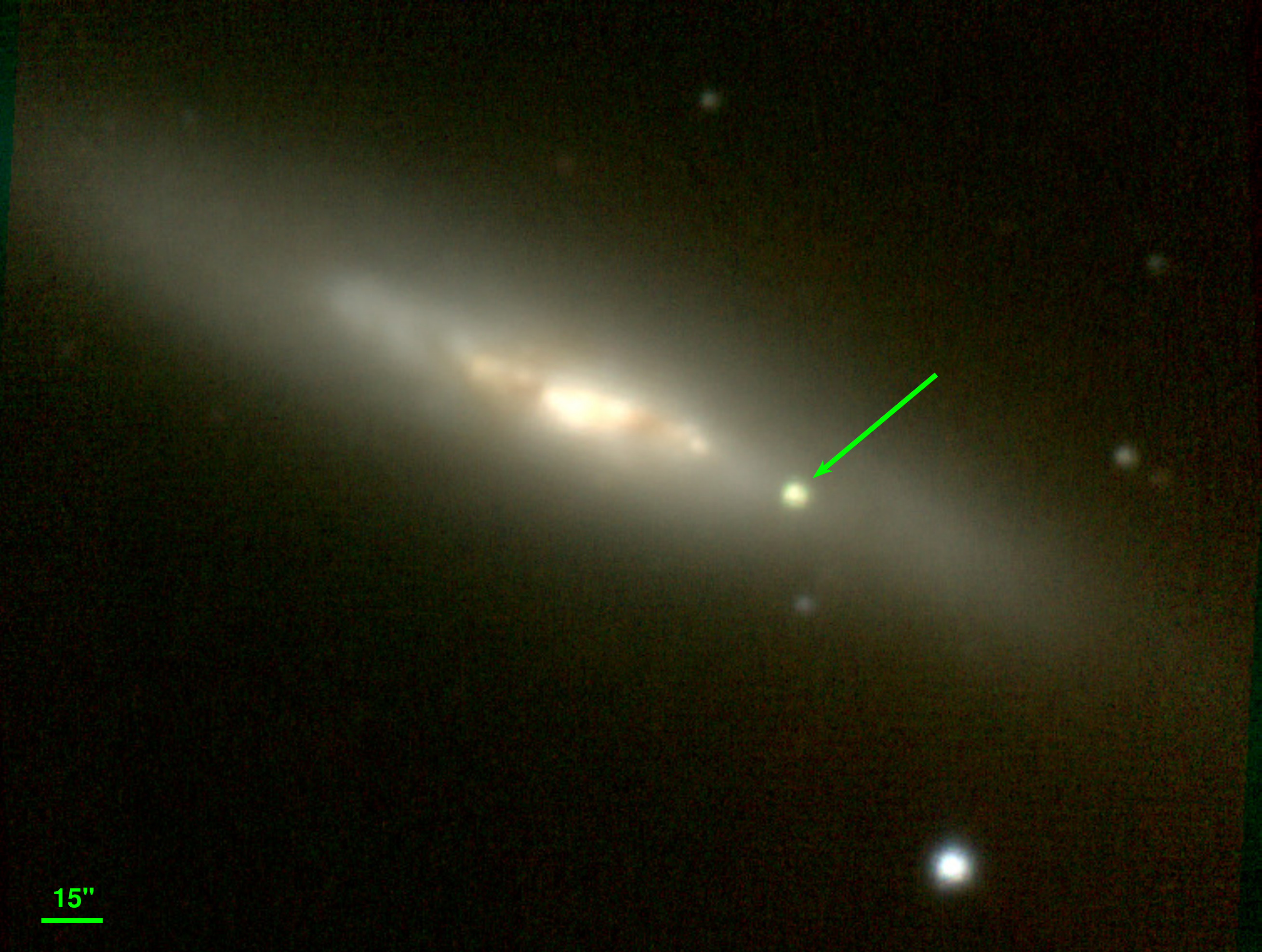}
\caption{Three color image of the SN 2014J field obtained with FLITECAM. Blue is J, green is H, and red is K. The arrow identifies SN 2014J, which is readily apparent 
as the point source SW of the nucleus and in the plane of the disk of M82. North is up and East is left.}
\end{center}
\label{Fig1}
\end{figure}

\begin{figure}
\begin{center}
\includegraphics[width=6.0in,height=8.0in]{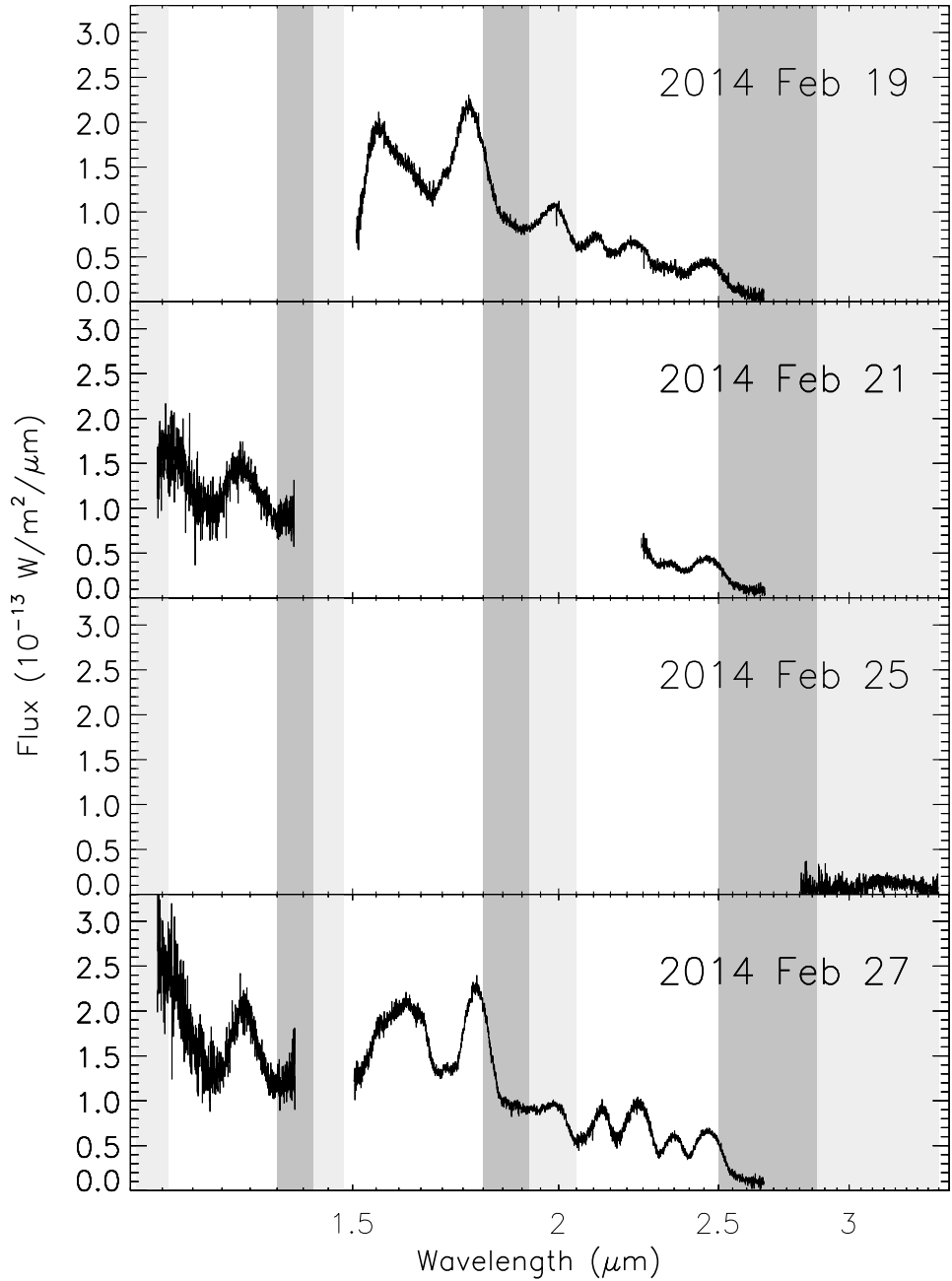}
\caption{FLITECAM flux-calibrated spectra of SN 2014J on the four different observing dates. Regions of strong telluric absorption 
from the ground (transmission $<20$\%) are shown in dark gray, while regions of moderate telluric absorption from the ground (transmission $<80$\%) are shown in light gray (cf., for example, Rayner et al. 2009).}
\end{center}
\label{Fig2}
\end{figure}

\begin{figure}
\begin{center}
\includegraphics[height=7.0in,angle=90]{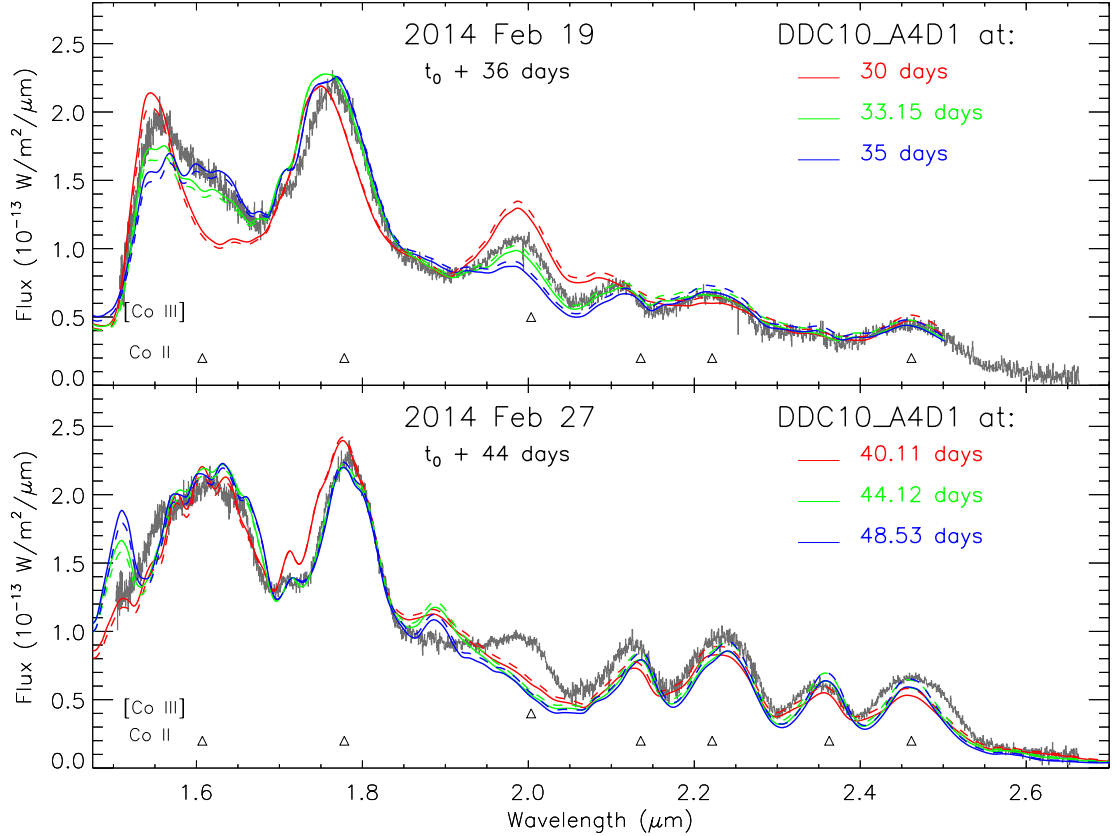}
\caption{Comparison between the observed $H+K$ band spectrum of SN 2014J obtained with FLITECAM (in gray) on 2014 Feb 19 UT (top)
and 2014 Feb.\ 27 UT (bottom) and the CMFGEN model DDC10 A4D1 of Dessart et al. (2014) for three different times past explosion. The model spectra have been scaled to match the flux levels of the data. Dashed lines are model spectra that have been reddened due to the effects of dust in both the Milky Way and M82. Triangles denote the locations of lines of Co II and [Co III] predicted to be strong in the model. }
\end{center}
\label{Fig3}
\end{figure}

\begin{figure}
\begin{center}
\includegraphics[height=7.5in,angle=90]{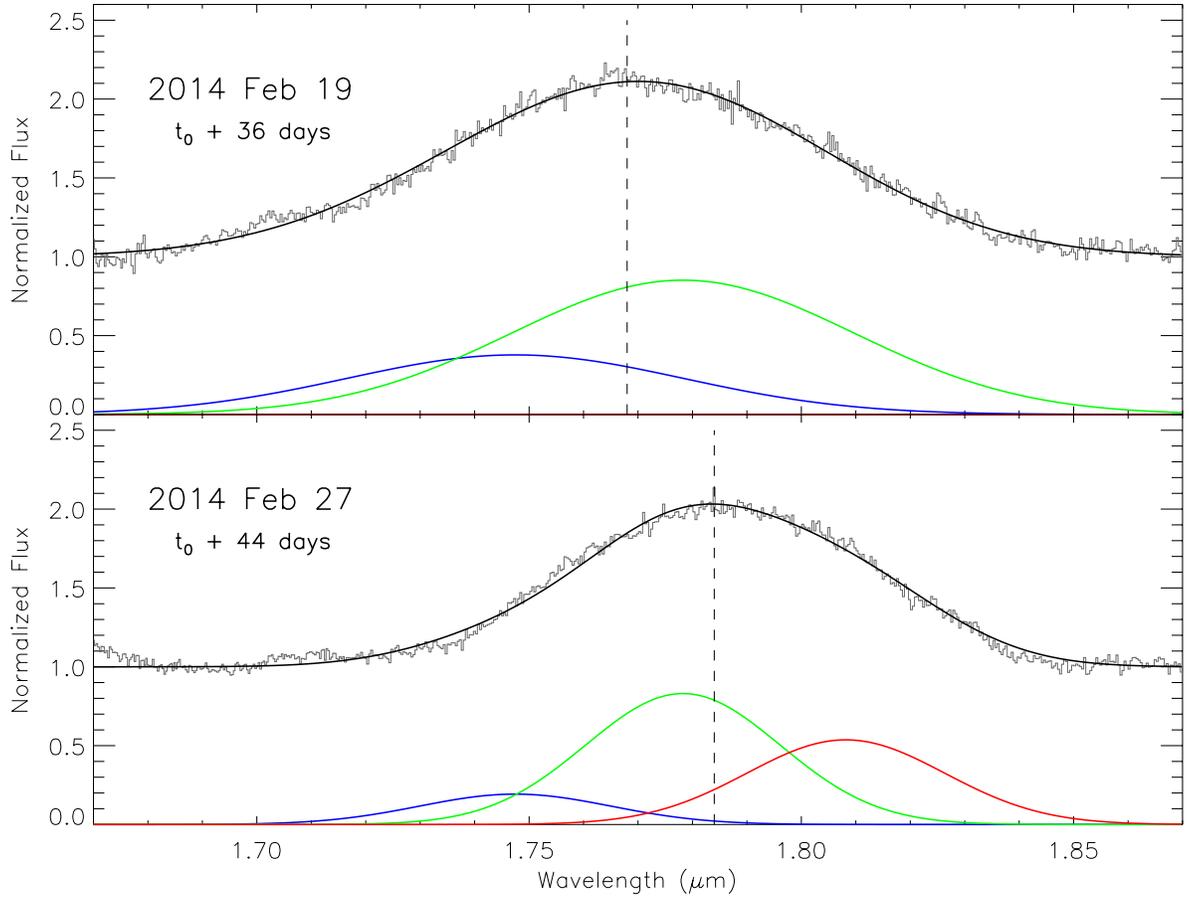}
\caption{Decomposition of the observed $1.77~\mu$m feature  in the spectra obtained on 2014 Feb.\ 19 (top) and 2014 Feb.\ 27 (bottom) into three Gaussians fixed at the locations of the strongest Co II lines 
($1.7463~\mu$m - blue, $1.7770~\mu$m - green, and $1.8069~\mu$m - red). The solid black is the sum. The dashed lines mark the centroids of the observed feature. The observed shift in the feature is due to 
the changing line strengths of the three components.}
\end{center}
\label{Fig4}
\end{figure}

\begin{figure}
\begin{center}
\includegraphics[height=7.0in,angle=90]{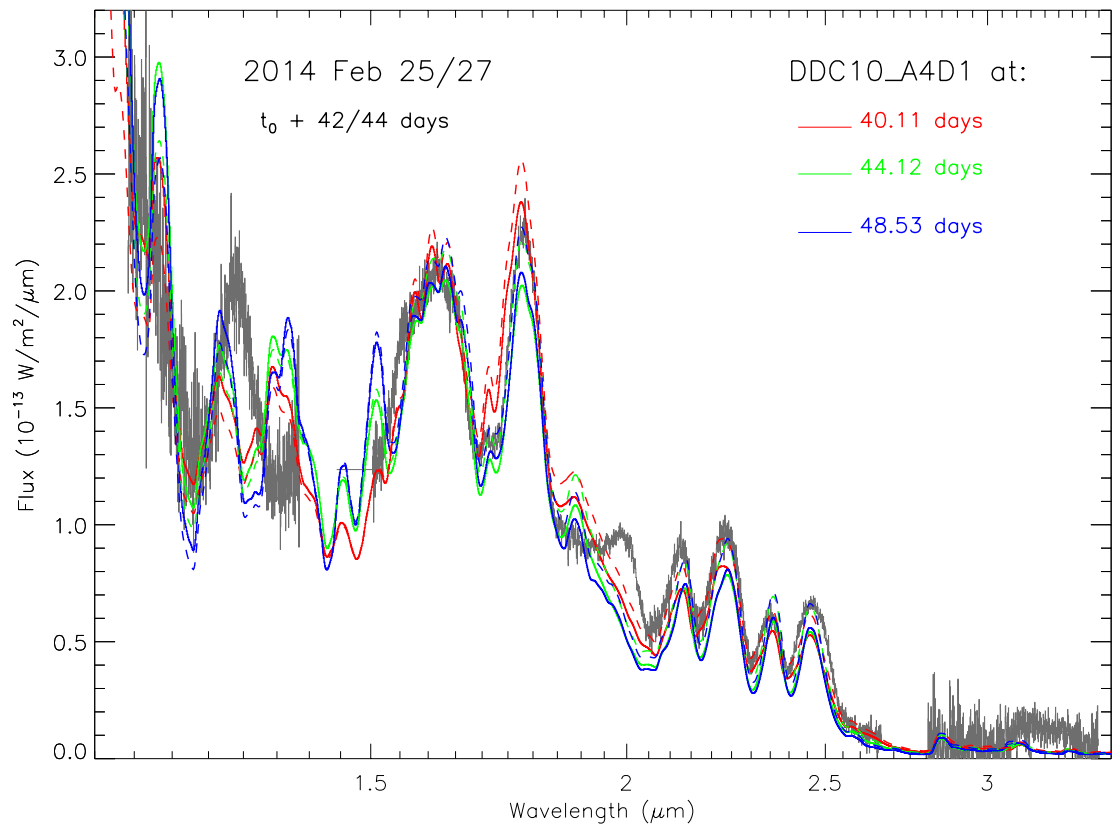}
\caption{Comparison between the observed spectrum of SN 2014J obtained with FLITECAM on 2014 Feb.\ 25 ($2.8-3.4~\mu$m) and Feb.\ 27 (JHK bands) UT (in gray) and the CMFGEN model DDC10 A4D1 
of Dessart et al. (2014) for three different times after explosion. The model spectra have been scaled to match the flux levels of the data. Dashed lines are model spectra that include reddening due to both Milky Way and M82 dust.}
\end{center}
\label{Fig5}
\end{figure}

\begin{figure}
\begin{center}
\includegraphics[height=7.0in,angle=90]{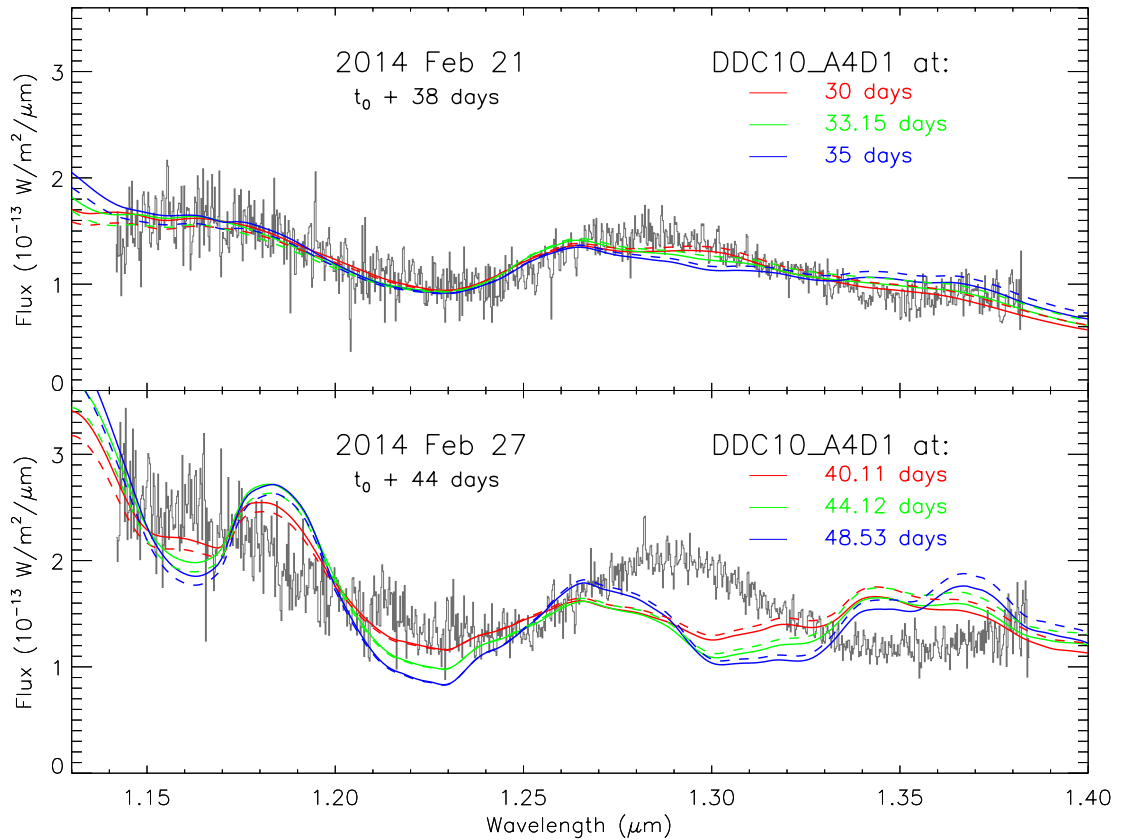}
\caption{Comparison between the observed $J$-band spectrum of SN 2014J obtained with FLITECAM on 2014 Feb.\ 21 UT (top)
and 2014 Feb.\ 27 UT (bottom) and the CMFGEN model DDC10 A4D1 of Dessart et al. (2014) for three different times past explosion. The model spectra have been scaled to match the flux levels of the data in this wavelength range. Dashed lines are model spectra that include reddening due to both Milky Way and M82 dust.}
\end{center}
\label{Fig6}
\end{figure}

\begin{figure}
\begin{center}
\includegraphics[height=7.0in,angle=90]{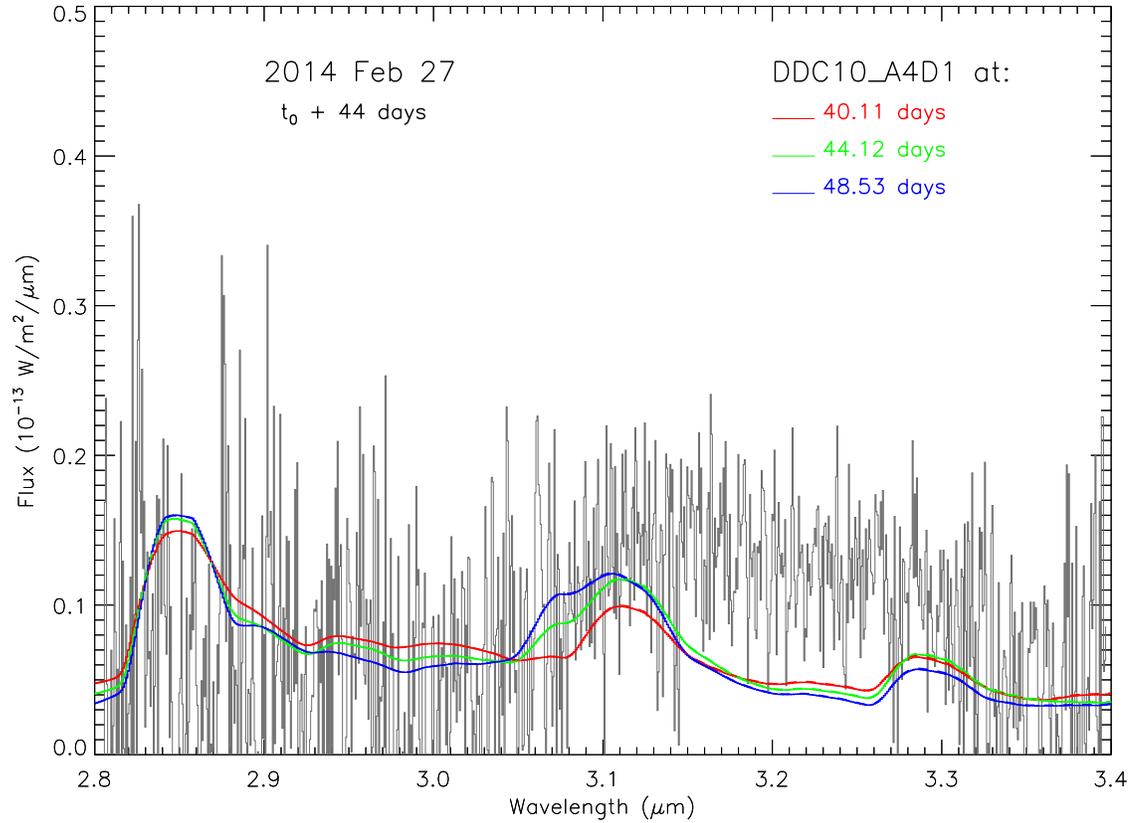}
\caption{Comparison between the observed $2.8 - 3.4~\mu$m spectrum of SN 2014J obtained with FLITECAM on 2014 Feb.\ 27 UT  and the CMFGEN model DDC10 A4D1 of Dessart et al. (2014) for 
three different times past explosion. The model spectra have been scaled to match the flux level of the data in this wavelength range.}
\end{center}
\label{Fig7}
\end{figure}

\begin{deluxetable}{l c c c c}
\tablewidth{0pt}
\tablecaption{FLITECAM/SOFIA Photometric Observations of SN 2014J}
\tablehead{
\colhead{MJD} & \colhead{UT Date} & Phase & Filter & Mag  \\
                         &                                & (days) &         &  \\
}
\startdata

56707.3            & 2014 02 19 07:11 &   17.5            & J      &  $10.97 \pm 0.06 $     \\
56707.3            & 2014 02 19 07:19 &   17.5            & H     &   $9.64 \pm 0.03$      \\
56707.3            & 2014 02 19 07:29 &   17.5           & K     &    $ 9.54  \pm 0.04 $\\
\hline
56709.3		& 2014 02 21 07:22 &   19.5            & J     & $10.93 \pm 0.05$      \\
56709.3		& 2014 02 21 07:32 &   19.5           & H     &     $9.56 \pm 0.04$   \\
56709.3		& 2014 02 21 07:43 &   19.5            & K      &  $ 9.49 \pm 0.04  $   \\        

\enddata
\tablecomments{Phase is the number of days past the time of maximum $B$ light (MJD 56689.8; Marion et al.\ 2014). }
\end{deluxetable}

\begin{deluxetable}{l c c c c c c c c c}
\tablewidth{0pt}
\tablecaption{Log of FLITECAM/SOFIA Spectroscopic Observations of SN 2014J}
\tablehead{
\colhead{MJD} & \colhead{UT Date} & $t-t_0$ & $t-t_{Bmax}$ & \colhead{Grism} & \colhead{Wavelengths} & \colhead{$t_{\rm exp}$} & \colhead{$N_{\rm exp}$}& Altitude & ZA \\
                         &                              &  (days)  &  (days)            & \colhead{ }  &  \colhead{(microns)}    & \colhead{(s)}           &                                      &  (feet)  & \\
}
\startdata

56707.4 & 2014 Feb 19 08:52 & 35.6 & 17.6 &  H\_C  & $1.500 - 1.718$ & 300 & 4 & 41,000 & 41 \\
56707.4 & 2014 Feb 19 10:26 & 35.7 & 17.6 & Hw\_B & $1.675 - 2.053$ & 150 & 8 & 41,000 & 37 \\
56707.5 & 2014 Feb 19 10:50 & 35.7 & 17.7 & Kw\_C & $1.910 - 2.276$ & 150 & 8 & 41,000 & 41 \\
56707.5 & 2014 Feb 19 11:28 & 35.7 & 17.7 & Kl\_A & $2.270 - 2.722$ & 180 & 2 & 41,000 & 43 \\
56707.5 & 2014 Feb 19 11:35 & 35.7 & 17.7 & Kl\_A & $2.270 - 2.722$ & 120 & 2 & 41,000 & 43 \\
\hline
56709.4 & 2014 Feb 21 09:35 & 37.7 & 19.6 & J\_B  & $1.141 - 1.385$ & 180 & 4 & 43,000 & 40 \\
56709.4 & 2014 Feb 21 09:53 & 37.7 & 19.6 & Kl\_A & $2.270 - 2.276$ & 120 & 6 & 43,000 & 40 \\
\hline
56713.4 & 2014 Feb 25 10:17 & 41.7 & 23.6 & LM\_C & $2.779 - 3.399$ & 150 & 4 & 40,000 & 35 \\
56713.4 & 2014 Feb 25 10:31 & 41.7 & 23.6 & LM\_C & $2.779 - 3.399$ & 200 & 12 & 40,000 & 35\\
\hline
56715.3 & 2014 Feb 27 06:20 & 43.5 & 25.5 & J\_B  & $1.141 - 1.385$ & 180 & 4 & 38,000 & 34 \\
56715.3 & 2014 Feb 27 06:37 & 43.5 & 25.5 & H\_C  & $1.500 - 1.718$ & 180 & 4 & 38,000 & 33 \\
56715.3 & 2014 Feb 27 06:53 & 43.5 & 25.5 & Hw\_B & $1.675 - 2.053$ & 180 & 4 & 38,000 & 33 \\
56715.3 & 2014 Feb 27 07:10 & 43.6 & 25.5 & Kl\_A & $2.270 - 2.722$ & 180 & 4 & 38,000 & 33 \\
56715.3 & 2014 Feb 27 07:27 & 43.6 & 25.5 & Kw\_C & $1.910 - 2.276$ & 180 & 4 & 38,000 & 33\\
\enddata
\label{abslines}
\tablecomments{UT date and time given are at the beginning of an exposure 
set. The exposure time $t_{\rm exp}$ is the time for a single exposure, and $N_{\rm exp}$ is the number of exposures obtained. $t-t_0$ is the number of days past $t_0$, the first light date estimated by Zheng et al.\ (2014); $t-t_{Bmax}$ is the number of days past maximum $B$ light estimated by Marion et al.\ (2014). ZA is the average zenith angle of the target during the observations.}
\end{deluxetable}

\end{document}